\documentclass[14pt]{revtex4}

\usepackage{graphicx}

\begin{document}

\title{A study of phantom scalar field cosmology using Lie and Noether symmetries}

\author{Sourav Dutta\footnote {sduttaju@gmail.com}}
\author{Subenoy Chakraborty\footnote {schakraborty.math@gmail.com}}
\affiliation{Department of Mathematics, Jadavpur University, Kolkata-700032, West Bengal, India.}

%%%%%%%%%%%%%%%%%%%%%%%%%%%%%%%%%%%%%%%%%%%%%%%%%%%%%%%%%%%%%%%%%%%%%%%%%%%%%%%%%%%%%%%%%%%%%%%%%%%%%%%%%%%%%%%%%%%%%%%%%%%%
\begin{abstract}
The paper deals with phantom scalar field cosmology in Einstein gravity. At first using Lie symmetry, the coupling function to the kinetic term and the potential function of the scalar field and the equation of state parameter of the matter field are determined and a simple solution is obtained. Subsequently, Noether symmetry is imposed on the Lagrangian of  the system. The symmetry vector is obtained and the potential takes a very general form  from which potential using Lie Symmetry can be obtained as a particular case. Then we choose a coordinate transformation $(a,\phi)\rightarrow(u,v)$ such that one of the transformed variables (say u) is a cyclic for the Lagrangian. Using conserved charge (corresponding to the cyclic coordinate) and the constant of motion, solutions are obtained.\\


\end{abstract}

\maketitle

\section{INTRODUCTION}

The observational evidences [1--7] that we have been getting since the fag end of the last century  are nicely accommodated in the frame work of standard cosmology by introducing an exotic matter [8--10] having large negative pressure-- the dark energy (DE). The natural and simplest candidate for DE is the cosmological constant [11--16]. But it is not very popular due to its extreme fine tuning and coincidence problem [17]. As a result, dynamical DE having (negative) variable equation of state [18] are widely used to show recent accelerated expansion. Subsequently, the idea of DE is extended further to consider matter with the equation of state parameter $w<-1$, which is not ruled out by observational data, and moreover, necessary to describe the current acceleration of the universe. Such a model is known as Phantom [19] DE model whose various aspects have been studied in [20--23]. However, inspite of lot of works (for reviews see references [8--10], [14-- 16], and [24]), still we are completely in the dark about the source of the DE, it is only characterized by its negative pressure. Although the $\Lambda$CDM model suggests the acceleration as a final stage but there is no priori reason why the acceleration of the universe will be in the final stage. On the other hand, other DE models may have different fate of the universe, for example, in phantom DE model the energy density may increase with time while the Hubble parameter and the curvature diverges in finite time and the final fate of the universe is a big-rip singularity [20, 21, 25]. It is puzzling [26] to have a sudden feature singularity for FRW cosmological models. Such singularity of pressure with finite energy density has some analogy to singularities in inhomogeneous models of the universe [27, 28].\\

In phantom cosmology, it has been shown recently that there is a nice duality [29] between phantom and ordinary matter fields similar to superstring cosmologies [30, 31]. Finally, due to increase of energy density in phantom cosmology, it is possible to connect the phantom driven (low energy mevscale) dark energy phase to the (high energy grand unified theories scale) inflationary era [32]. In the present work, a phantom model of the universe is considered for which gravity is coupled to a scalar field with a scalar coupling function and a potential. Basic geometrical symmetries (connected to the space-time) particularly, Lie point and Noether are very useful in physical problems. In fact Noether symmetries are usually related with some conserved quantities (termed as conserved charges) which can be chosen as a selection criteria to discriminate similar physical processes for example several DE models [33--40]. Mathematically, Lie point/Noether symmetries play a crucial role in any physical problem, as they provide first/Noether integrals which can be used to simplify a given system of differential equation or to determine the integrability of the system. Further, Noether symmetries are also used to examine self-consistency of phenomenological physical models. Moreover, imposing symmetry to a physical problem, one may be able to restrict physical parameters like equation of state of a fluid [41], potential of a scalar field etc. Recently, an explicit study of physical systems moving in a Riemannian space and the corresponding Lie point/Noether symmetries are discussed in [42--48].\\

                    The motivation of the present work is to study the phantom scalar field cosmology in the perspective of symmetry. Our aim is two fold:\\
										
										(a) Specifying the coupling function and the scalar potential from geometric principles.\\ 
										
										(b)Cosmological solutions using the symmetries.\\

			 Geometric approach have been applied to symmetries in the minisuperspace approach and they are related with basic geometric properties of the minisuperace, hence there are geometric selection rules and that have been studied in  different gravitational theories [54, 55, 56] \\

										The paper is organized as follows: Section II deals with basic equations for the phantom scalar field model while Lie point symmetry has been imposed in Section III. Section IV deals with the Noether symmetry approach and the corresponding solutions. At the end, symmetry of the work is presented in section V. 
\section{Basic Equation And Autonomous System}
The action of a scalar field $\phi$ coupled to gravity with coupling parameter, a function of the scalar field $\phi$ has the form [49, 54] :
\begin{equation}
A=\int d^4x\sqrt-g[\frac{R}{2k^2}-\frac{1}{2}\lambda(\phi)g^{\mu\nu}({\Delta_\mu}\phi)({\Delta_\nu}\phi)-V(\phi)]+S_m,
\end{equation}
where $k^2=8\pi$G is the gravitational coupling, $\lambda(\phi)$ the coupling parameter is an arbitrary function of $\phi$, $V(\phi)$ is the potential of the scalar field and $S_m$ is the action of the matter field which is chosen as the cold dark matter (DM) in the form of perfect fluid with constant equation of state. For flat FRW model, if we vary the action with respect to the metric variable then we have the Einstein equations :
\begin{equation}
3\frac{\dot{a}^2}{a^2}=\kappa^2\left(\rho_m+\frac{\lambda(\phi)}{2}\dot{\phi}^2+V(\phi)\right),
\end{equation}
\begin{equation}
2\frac{\ddot{a}}{a^2}=-\frac{\kappa^2}{3}\left[(3\gamma-2)\rho_m+2(\lambda(\phi)\dot{\phi}^2-V(\phi))\right],
\end{equation}
with $\rho_{\phi}=\frac{1}{2}\lambda(\phi)\dot{\phi}^2+V(\phi)$, and $p_{\phi}=\frac{1}{2}\lambda(\phi)\dot{\phi}^2-V(\phi)$, the energy density and thermodynamic pressure of the scalar field. Assuming there is no interaction between the scalar field and cold DM, then the energy conservation relations take the form :
\begin{equation}
\dot{\rho_m}+3\gamma\rho_m=0,~~~ i.e.,~~~~ \rho_m=\rho_0a^{-3\gamma},
\end{equation}
and
\begin{equation}
\dot{\rho_\phi}+3H(\rho_\phi+p_\phi)=0.
\end{equation}
As a result, the evolution of the scalar field is given by
\begin{equation}
\lambda(\phi)\ddot{\phi}+\frac{1}{2}\lambda^\prime(\phi)\dot{\phi}^2+3H\lambda(\phi)\dot{\phi}+\frac{\partial V}{\partial\phi}=0.
\end{equation}
Due to non-linearity in nature and complicated in form, the evolution equations (i.e., eq. (3) and eq. (6)) have been studied [49] through qualitative analysis. Using the auxiliary variables\\

$~~~~~~~~~~~~~~~~~~~~~~~~~~~~~~~~~~~~~~~~~~~~~~x=\frac{\sqrt{\lambda(\phi)}}{\sqrt{\epsilon}H}\dot{\phi},~~y=\frac{\sqrt{V(\phi)}}{\sqrt{3}H},~~z=\frac{\sqrt{6}}{\phi}$,\\

the evolution equations can be written as autonomous system\\

$$x^\prime=\frac{3}{2}x(x^2-y^2-1)-\alpha_1(z)y^2,$$
$$y^\prime=y\left[\alpha_1(z)x+\frac{3}{2}(x^2-y^2+1)\right],$$
and\\
$$z^\prime=-\frac{xz^2}{\sqrt{\lambda_1(z)}},$$\\

where $\alpha_1(z)=\alpha(\phi)=\sqrt{\frac{3}{2}}\frac{V^\prime(\phi)}{V(\phi)\sqrt{\lambda(\phi)}},~~\lambda_1(z)=\lambda(\phi),$~~ and $'{\prime}~'$ stands for differentiation with respect to $N=ln~a$.\\

Note that the auxiliary variables are not independent rather they are connected by the relation\\

$$x^2+y^2+\Omega_m=1.$$\\

In Ref [49] the potential $V(\phi)$ and the coupling function $\lambda(\phi)$ has been chosen in an ad hoc manner as exponential or power law form. From the stability analysis it is generally, concluded that although there are late time attractors with phantom scalar field but they are not quantum mechanically stable.
\section{Lie Point Symmetry and Consequences}
Suppose we have a system of second order ordinary differential equations of the form
\begin{equation}
\ddot{x}^i=w^i(t,x^j,\dot{x}^j).
\end{equation}
Now a vector field $X=\xi\frac{\partial}{\partial t}+\eta^i\frac{\partial}{\partial x^i}$ in the augmented space ($t,x^i$) is the generator of a Lie point symmetry of the above system of ordinary differential equations if the following relation hold [44] :
\begin{equation}
X^{[2]}\left(\ddot{x}^i-w(t,x^i,\dot{x}^i)\right)=0,
\end{equation}
where $X^{[2]}$ stands for the second prolongation of X and has the expression
\begin{equation}
x^{[2]}=\xi\partial_t+\eta^i\partial_i+(\dot{\eta}^i-\dot{x}^i\dot{\xi})\partial\dot{x}^i+(\ddot{\eta}^i-\dot{x}^i\ddot{\xi}-2\ddot{x}^i\dot{\xi})\partial\ddot{x}^i.
\end{equation}
Equivalently, the lie point symmetry condition can be written as 
\begin{equation}
\left[X^{[1]},A\right]=\lambda(x^i)A,
\end{equation}
where, 
\begin{equation}
x^{[1]}=\xi\partial_t+\eta^i\partial_i+(\dot{\eta}^i-\dot{x}^i\dot{\xi})\partial\dot{x}^i,
\end{equation}
is the first prolongation vector of X and the Hamiltonian vector field $A$ has the expression
\begin{equation}
A=\partial_t+\dot{x}^i\partial_{x^i}+w^i(t,x^j,\dot{x}^j)\partial_{\dot{x}^i}.
\end{equation}
In the present phantom cosmology, we have 3D augmented space ($t,a,\phi$) and the evolution equations(2), (3) and (6) can be written as
\begin{equation}
\ddot{a}=\frac{\dot{a}^2}{a}-\frac{\kappa^2}{2}a\left(\gamma E_ma^{-3\gamma}+\lambda(\phi)\dot{\phi}^2\right)\equiv w_1(t,a,\dot{a},\phi,\dot{\phi}),
\end{equation}
and
\begin{equation}
\ddot{\phi}=-\frac{1}{2}\frac{\lambda^\prime(\phi)}{\lambda(\phi)}\dot{\phi}^2-3H\dot{\phi}-\frac{1}{\lambda(\phi)}\frac{dV}{d\phi}\equiv w_2(t,a,\dot{a},\phi,\dot{\phi}).
\end{equation}
So the infinitesimal generator of the point transformation corresponding to equations (13) and (14) can be written as
\begin{equation}
X=\xi\frac{\partial}{\partial t}+\eta_1\frac{\partial}{\partial a}+\eta_2\frac{\partial}{\partial \phi}+\eta_1^\prime\frac{\partial}{\partial \dot{a}}+\eta_2^\prime\frac{\partial}{\partial \dot{\phi}},
\end{equation}
where $\xi=\xi(t,a,\phi)$  and $\eta_i=\eta_i(t,a,\phi)$,$i=1,2.$
Using the total derivative operator as
\begin{equation}
\frac{d}{dt}=\frac{\partial}{\partial t}+\dot{a}\frac{\partial}{\partial a}+\dot{\phi}\frac{\partial}{\partial \phi} ,
\end{equation}
one can write
\begin{equation}
\eta_1^\prime=\frac{d\eta_1}{dt}-\dot{a}\frac{d\xi}{dt}~~~~~~and~~~~~ \eta_2^\prime=\frac{d\eta_2}{dt}-\dot{\phi}\frac{d\xi}{dt}.
\end{equation}
Now, the lie point symmetry condition corresponding to equation (13) can be written as [44]
\begin{equation}
Xw_1=\eta_1^{\prime\prime},
\end{equation}
and that for equation (14) is
\begin{equation}
Xw_2=\eta_2^{\prime\prime},
\end{equation}
with,
$$\eta_1^{\prime\prime}=\frac{d\eta_1^{\prime}}{dt}-\ddot{a}\frac{d\xi}{dt}=\frac{d\eta_1^{\prime}}{dt}-w_1\frac{d\xi}{dt},$$
\begin{equation}
\eta_2^{\prime\prime}=\frac{d\eta_2^{\prime}}{dt}-\ddot{\phi}\frac{d\xi}{dt}=\frac{d\eta_2^{\prime}}{dt}-w_2\frac{d\xi}{dt}.
\end{equation}
Solving the overdetermined system of equations corresponding to the symmetry conditions (18) and (19), the component of the symmetries takes the form:
$$\xi=\alpha t+\beta,~~~~~~~\eta_1=\alpha a,~~~~~~~\eta_2=\alpha\phi,$$with
\begin{equation}
\lambda=\frac{\lambda_0}{\phi^2}~~~~,~~~V(\phi)=V_0-\frac{\mu_0}{\phi^2},
\end{equation}
and the equation of state parameter $\gamma$ takes values $0$ and $\frac{2}{3}$. Here, $\alpha,\beta,\lambda_0,V_0~~and~~\mu_0$ are arbitrary constants. Hence the symmetry vector is
\begin{equation}
X=(\alpha t+\beta)\partial_t+\alpha a\frac{\partial}{\partial a}+\alpha\phi\frac{\partial}{\partial \phi}.
\end{equation}
The characteristic equation corresponding to this symmetry is given by
\begin{equation}
\frac{dt}{\alpha t+\beta}=\frac{da}{\alpha a}=\frac{d\phi}{\alpha\phi},
\end{equation}
which gives an one parameter family of functions
\begin{equation}
a=t+\frac{\beta}{\alpha}=\phi.
\end{equation}
Now using the above one parameter family of functions to the field equations(2), (3), and (6) the constants are related by the relations:
$$\lambda_0=\frac{2}{\kappa^2}=-\mu_0~~~~~,~~~~~~V_0=-\rho_0,~~~~~~~~~~\textmd{when }\gamma=0,$$
\begin{equation}
\mu_0=\frac{2}{3}\left(\rho_0-\frac{3}{\kappa^2}\right)=-\lambda_0~~~,~~~V_0=0,~~~~~\textmd{when }\gamma=\frac{2}{3}.
\end{equation}
\section{The Noether Symmetry Approach}
In Noether symmetry approach, some conserved quantities corresponding to a dynamical system can be determined from the invariance of the corresponding Lagrangian under the lie derivative along an appropriate vector field. Thus due to this symmetry constraints are imposed on dynamics and as a result it is possible to solve the equations of motion. In general, for a point-like canonical Lagrangian L which depends on the variables $q^\alpha(x^i)$ and on their derivatives $\partial_jq^\alpha(x^i)$, the Euler-Lagrange equations are\\
\begin{equation}
\partial_k\left(\frac{\partial L}{\partial \partial_j q^\alpha}\right)=\frac{\partial L}{\partial q^\alpha}.
\end{equation}

Now contracting with some unknown functions $\lambda^\alpha(q^\beta)$, we have\\
\begin{equation}
\lambda^\alpha\left[\partial_k\left(\frac{\partial L}{\partial \partial_j q^\alpha}\right)-\frac{\partial L}{\partial q^\alpha}\right]=0,
\end{equation}

which on simplification can be written as\\
\begin{equation}
\mathcal{L}_XL=\partial_k\left(\lambda^\alpha\frac{\partial L}{\partial \partial_j q^\alpha}\right).
\end{equation}

Here $\mathcal{L}_X$ stands for the Lie derivative along the vector field\\
\begin{equation}
X=\lambda^\alpha\frac{\partial}{\partial q^\alpha}+\left(\partial_j\lambda^\alpha\right)\frac{\partial}{\partial \partial_j q^\alpha}.
\end{equation}

Thus as a consequence of Noether theorem (which states that $\mathcal{L}_XL=0$), we can say that the Lagrangian $L$ is invariant along the vector field $X$ which is also called the generator of symmetry. Consequently, we can define the charge [51]

\begin{equation}
J^i=\lambda^\alpha\frac{\partial L}{\partial \partial_i q^\alpha},
\end{equation}

which is conserved as\\
\begin{equation}
\partial_iJ^i=0,
\end{equation}

The energy function associated with Lagrangian is
\begin{equation}
E=\frac{\partial \mathcal{L}}{\partial \dot{q}^\alpha}\dot{q}^\alpha-L,
\end{equation}

which is the total energy $(T+ V)$ of the system and is a constants of motion [35].\\

Due to the presence of the Noether symmetry, evolution equations are reduced and hence the dynamics can be solved exactly. Further, if the symmetry (i.e., conserved quantity) have some physical meaning, then Noether symmetry approach can be used to select reliable models [37]. \\

In the present cosmological model with phantom scalar field, the point like Lagrangian takes the form

\begin{equation}
L=3a\dot{a}^2+\kappa^2\rho_0a^{3(1-\gamma)}-\frac{\kappa^2}{2}a^3\lambda(\phi)\dot{\phi}^2+\kappa^2a^3V(\phi).
\end{equation}

Here, the infinitesimal generator of the Noether symmetry( We recall that we consider a special case of Noether's theorem. The application of the completes Noether's theorem in  Scalar field cosmology can be found in [54] ) is\\ 
\begin{equation}
X=\alpha\frac{\partial}{\partial a}+\beta\frac{\partial}{\partial \phi}+\dot{\alpha}\frac{\partial}{\partial \dot{a}}+\dot{\beta}\frac{\partial}{\partial \dot{\phi}},
\end{equation}

where $\alpha=\alpha(a,\phi),~~\beta=\beta(a,\phi),~~\dot{\alpha}=\frac{\partial \alpha}{\partial a}\dot{a}+\frac{\partial \alpha}{\partial \phi}\dot{\phi},~~and~~\dot{\beta}=\frac{\partial \beta}{\partial a}\dot{a}+\frac{\partial \beta}{\partial \phi}\dot{\phi}.$\\

Now, the existence of Noether symmetry of the above Lagrangian demands $\mathcal{L}_X L=0$ and hence $\alpha, \beta$ will satisfy the following partial differential equations :\\
\begin{equation}
\alpha+2a\frac{\partial \alpha}{\partial a}=0,
\end{equation}
\begin{equation}
6\frac{\partial \alpha}{\partial \phi}-\kappa^2a^2\lambda(\phi)\frac{\partial \beta}{\partial a}=0,
\end{equation}
\begin{equation}
3\alpha\lambda+\beta a\lambda^\prime+2a\lambda\frac{\partial \beta}{\partial \phi}=0,
\end{equation}
and
$$3\alpha\rho_0+3\alpha V(\phi)+\beta aV^\prime(\phi)=0,~~~~~~~~for~~~\gamma=0,$$
\begin{equation}
~~~~~~~~~~3\alpha V(\phi)+\beta aV^\prime(\phi)=0,~~~~~~~for~~~~\gamma=1.
\end{equation}

In order to solve this coupled partial differential equation, we use the method of separation of variable as
\begin{equation}
\alpha=A_1(a)B_1(\phi)~,~~~~~~~~\beta=A_2(a)B_2(\phi).
\end{equation}

At first we choose $\lambda(\phi)=\frac{\lambda_0}{\phi^2}$  from the Lie point symmetry and the possible solutions are the following:\\

{\bf Case I: $\lambda_0 > 0$}
\begin{equation}
\alpha=\frac{A_0}{\sqrt{a}}cosh(p~ln~\phi+b_1),
\end{equation}
\begin{equation}
\beta=-\frac{4A_0}{\kappa^2 \lambda_0}pa^{-\frac{3}{2}}\phi sin h(p~ln~\phi+b_1),
\end{equation}
\begin{equation}
V=\left\{
\begin{array}{ll}
V_0sinh^2(p~ln~\phi+b_1), & \mbox{ for } \gamma=1\\
V_0sinh^2(p~ln~\phi+b_1)-\rho_0, & \mbox{ for } \gamma=0
\end{array}
\right\}.
\end{equation}
Where,~~~$p^2=\frac{3}{8}\kappa^2\left|\lambda_0\right|$,~~$A_0$ is arbitrary.\\

{\bf Case II: $\lambda_0 < 0$}
\begin{equation}
\alpha=\frac{A_0}{\sqrt{a}}cos(p~ln~\phi+b_1),
\end{equation}
\begin{equation}
\beta=-\frac{4A_0}{\kappa^2 \left|\lambda_0\right|}pa^{-\frac{3}{2}}\phi sin(p~ln~\phi+b_1),
\end{equation}
\begin{equation}
V=\left\{
\begin{array}{ll}
V_0sin^2(p~ln~\phi+b_1), & \mbox{ for } \gamma=1\\
V_0sin^2(p~ln~\phi+b_1)-\rho_0, & \mbox{ for } \gamma=0
\end{array}
\right\}.
\end{equation}

It should be noted here that if we choose $\lambda=\lambda_0$, a constant then the infinitesimal generator of the Noether symmetry coincides with [38].\\

Now we shall assume the potential in the form of exponential function, i.e., $V=V_0e^{\mu \phi}$ where $V_0$ and $\mu$ are the constants. Using the set of 1st order Partial differential equations (35)--(38) we can determine the function  $\lambda(\phi)$ and the infinitesimal generator of the Noether symmetry as follows \\

$\underline{When~\gamma=0}$

\begin{eqnarray}
\alpha&=&\alpha_0a^{\frac{-1}{2}}\left[e^{\mu \phi}\left(1+\frac{\rho_0}{V_0}e^{-\mu \phi}\right)-\frac{\rho_0}{V_0}ln\left(e^{\mu \phi}+\frac{\rho_0}{V_0}\right)\right]^\frac{1}{2},
\nonumber
\\
\beta&=&-\beta_0a^\frac{-3}{2}\left(1+\frac{\rho_0}{V_0}e^{-\mu \phi}\right)\left[e^{\mu \phi}\left(1+\frac{\rho_0}{V_0}e^{-\mu \phi}\right)-\frac{\rho_0}{V_0}ln\left(e^{\mu \phi}+\frac{\rho_0}{V_0}\right)\right]^\frac{1}{2},
\nonumber
\\
\lambda&=&\lambda_0\frac{e^\mu \phi}{\left(1+\frac{\rho_0}{V_0}e^{-\mu \phi}\right)^2\left[e^{\mu \phi}\left(1+\frac{\rho_0}{V_0}e^{-\mu \phi}\right)-\frac{\rho_0}{V_0}ln\left(e^{\mu \phi}+\frac{\rho_0}{V_0}\right)\right]^\frac{1}{2}}.
\end{eqnarray}

$\underline{When~\gamma=1}$
\begin{eqnarray}
\alpha&=&\alpha_0a^\frac{-1}{2}\left[b_0+c_0e^{\mu \phi}\right]^\frac{1}{2},
\nonumber
\\
\beta&=&-\beta_0a^\frac{-3}{2}\left[b_0+c_0e^{\mu \phi}\right]^\frac{1}{2},
\nonumber
\\
\lambda&=&\lambda_0\frac{e^{\mu \phi}}{\left[b_0+c_0e^{\mu \phi}\right]},
\end{eqnarray}
where, $\alpha_0~,\beta_0~,\lambda_0~,b_0~,c_0$ are constants.\\

Hence the infinitesimal generator of the Noether symmetry is completely determined.\\

In the next step, we shall determine a change of variables in the augmented space, i.e., $(a,\phi)\rightarrow(u,v)$ such that one of the variables become cyclic, i.e., the transformed Lagrangian reduced the dynamical system and then it can be solved.\\

The explicit form of the two constant of motion namely conserved charge (see eq. (30)) and conserved energy (see eq. (32)) are given by:
\begin{equation}
~~~~J^\kappa=\alpha\frac{\partial L}{\partial \dot{a}}+\beta \frac{\partial L}{\partial \dot{\phi}},
\end{equation}
and
\begin{equation}
E=\dot{a}\frac{\partial L}{\partial \dot{a}}+\dot{\phi} \frac{\partial L}{\partial \dot{\phi}}-L.
\end{equation}
Further introducing the carton one form [35] namely
\begin{equation}
\theta_L=\frac{\partial L}{\partial \dot{a}}da+\frac{\partial L}{\partial \dot{\phi}}d\phi.
\end{equation}

The conserved charge can be written as

\begin{equation}
J^\kappa=i_X\theta_L,
\end{equation}

where, $i_X\theta_L$ indicates contractions between the vector field $X$ and the differential form $\theta$.\\

So by a point transformation $(a,\phi)\rightarrow(u,v)$ the vector field $X$ becomes
\begin{equation}
\widetilde{X}=(i_X du)\frac{\partial}{\partial u}+(i_xdv)\frac{\partial}{\partial v}+\left(\frac{d}{dt}(i_Xdu)\right)\frac{d}{d\dot{u}}+\left(\frac{d}{dt}(i_xdv)\right)\frac{d}{d\dot{v}}  .
\end{equation}

Note that $\widetilde{X}$ is still the lift of a vector field defined on the space of positions. As $X$ is a symmetry so we choose a coordinate transformation such that

\begin{equation}
i_X du=1~,~~~~~~~~~~~~~~~~~~~i_X dv=0,
\end{equation}

so that  $\widetilde{X}=\frac{\partial}{\partial u}~~~~and~~~~\frac{\partial L}{\partial u}=0$. Thus $u$ is a cyclic coordinate and the dynamics can be reduced [38].\\

Hence by solving the equations (53) $a$ and $\phi$ can be obtained as\\

{\bf for: $\lambda_0 > 0$}
$$a=\left(\frac{3A_0}{2}\right)^{\frac{2}{3}}\left(u^2-v^2\right)^{\frac{1}{3}},$$
\begin{equation}
\phi=exp\left[\frac{1}{p}\left(tanh^{-1}(\frac{v}{u})-b_1\right)\right].
\end{equation}

{\bf for: $\lambda_0 < 0$}
$$a=\left(\frac{3A_0}{2}\right)^{\frac{2}{3}}\left(u^2+v^2\right)^{\frac{1}{3}},$$
\begin{equation}
\phi=exp\left[\frac{1}{p}\left(tan^{-1}(\frac{v}{u})-b_1\right)\right].
\end{equation}

The transformed Lagrangian takes the form:

\begin{equation}
L=\left\{
\begin{array}{llll}
3A_0^2(\dot{u}^2-\dot{v}^2)+\frac{9}{4}\kappa^2A_0^2V_0v^2+\kappa^2\rho_0 & \mbox{ for } \lambda_0{>}0,\gamma=1\\
3A_0^2(\dot{u}^2-\dot{v}^2)+\frac{9}{4}\kappa^2A_0^2V_0v^2 & \mbox{ for } \lambda_0{>}0,\gamma=0\\
3A_0^2(\dot{u}^2+\dot{v}^2)+\frac{9}{4}\kappa^2A_0^2V_0v^2+\kappa^2\rho_0 & \mbox{ for } \lambda_0{<}0,\gamma=1\\
3A_0^2(\dot{u}^2+\dot{v}^2)+\frac{9}{4}\kappa^2A_0^2V_0v^2 & \mbox{ for } \lambda_0{<}0,\gamma=0
\end{array}
\right\},
\end{equation}

with $u$ as a cyclic coordinate.\\

Now the conserved charge gives

\begin{equation}
J_u=\frac{\partial L}{\partial \dot{u}}=6A_0^2\dot{u}=J_0,
\end{equation}
which integrates to give, $u(t)=u_0t+u_1$.\\\\
Similarly, the conserved energy in terms of $u,v$ gives 
\begin{equation}
E_l=3A_0^2(\dot{u}^2-\epsilon\dot{v}^2)-\frac{9}{4}\kappa^2A_0^2V_0v^2-\kappa^2\rho_0\gamma=0,
\end{equation}
where $\epsilon=\pm1$ corresponding to $\lambda_0{>}or{<}0~~and~~~\gamma=0~~or~~1$.\\\\
Now using the solution for $u$, the solution of $v$ takes the form

\begin{equation}
v=\left\{
\begin{array}{llll}
\frac{\sqrt{u_0^2-\gamma^2m^2}}{l} sin[l(t-t_0)] & \mbox{ for } \epsilon=+1\\
\frac{\sqrt{m^2-u_0^2}}{l} cos[l(t-t_0)] & \mbox{ for } \epsilon=+1\\
\frac{\sqrt{u_0^2-\gamma^2m^2}}{l} cosh[l(t-t_0)] & \mbox{ for } u_0{>}m_0\\
\frac{\sqrt{m^2-u_0^2}}{l} sinh[l(t-t_0)] & \mbox{ for } m{>}u_0
\end{array}
\right\},
\end{equation}
where $\gamma=0~~~or~~~1$ and $l^2=\frac{3\kappa^2V_0}{4},~~~m^2=\frac{\kappa^2\rho_0}{3A_0^2}$.\\

Hence from (54) or (55) the explicit form of the scale factor and the phantom scalar field are given by\\

$\underline{A.~for~\lambda_0~{>}~0,~\gamma=1}$
\begin{equation}
a(t)=\left\{
\begin{array}{ll}
\left(\frac{3A_0}{2}\right)^{\frac{2}{3}}\left[u_0^2t^2-\frac{(u_0^2-m^2)}{l^2}sin^2(lt)\right]^{\frac{1}{3}}\\
\left(\frac{3A_0}{2}\right)^{\frac{2}{3}}\left[u_0^2t^2-\frac{(m^2-u_0^2)}{l^2}cos^2(lt)\right]^{\frac{1}{3}}
\end{array}
\right\},
\end{equation}
and
\begin{equation}
\phi(t)=\left\{
\begin{array}{ll}
exp\left[\frac{1}{p}(tanh^{-1}[\frac{\sqrt{u_0^2-m^2}}{lu_0t}sin(lt)]-b_1)\right]\\
exp\left[\frac{1}{p}(tanh^{-1}[\frac{\sqrt{m^2-u_0^2}}{lu_0t}cos(lt)]-b_1)\right]
\end{array}
\right\},
\end{equation}
according as $u_0~{>}~m~~or~~u_0~{<}~m$. In these solutions we have chosen $u_1=0=t_0$ without any loss of generality.\\

\begin{figure}
\begin{minipage}{0.4\textwidth}
\includegraphics[width=1.0\textwidth]{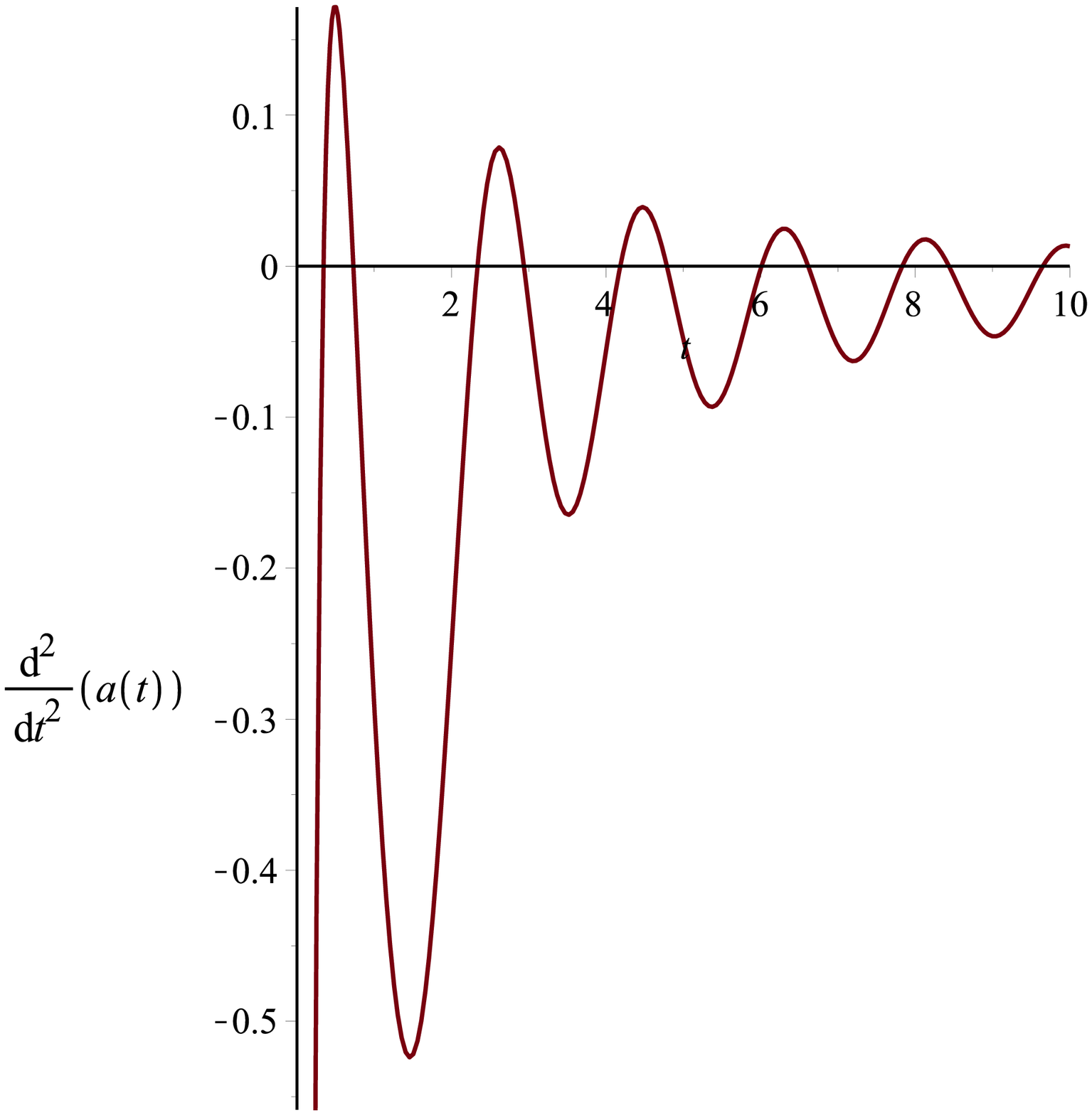}\\
Figure 1: represents $\ddot{a}(t)$ against $t$ for $\lambda_0>0,~\gamma=1$ with $A_0=1,~u_0=1,~l^2=3,~m^2=\frac{1}{3}$.
\end{minipage}
\begin{minipage}{0.4\textwidth}
\includegraphics[width=1.0\textwidth]{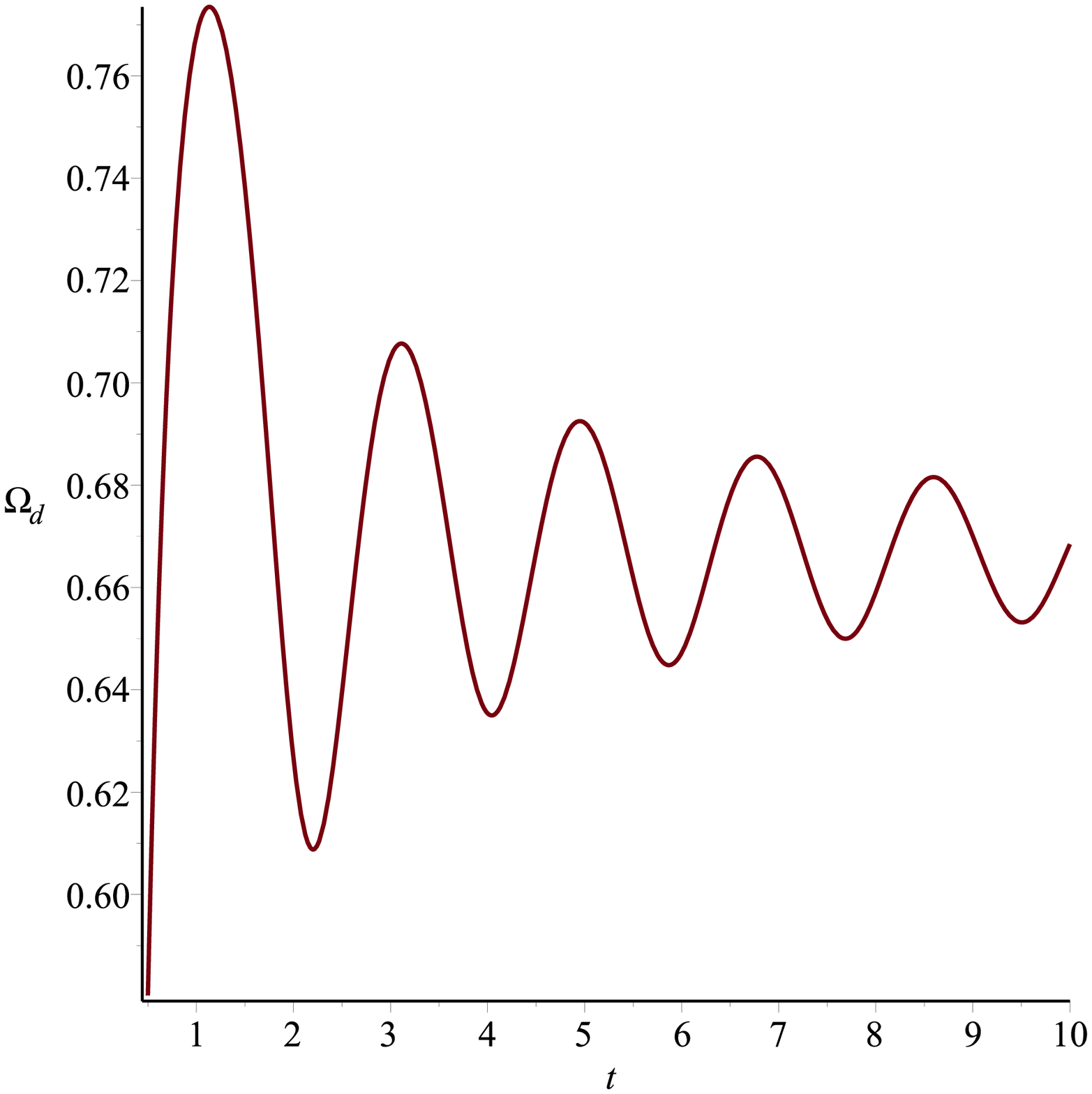}\\
Figure 2: shows the variation of  $\Omega_d$ against $t$ for $\lambda_0>0,~\gamma=1$ with $A_0=1,~u_0=1,~l^2=3,~m^2=\frac{1}{3}$.
\end{minipage}
\end{figure}
\begin{figure}
\begin{minipage}{0.4\textwidth}
\includegraphics[width=1.0\textwidth]{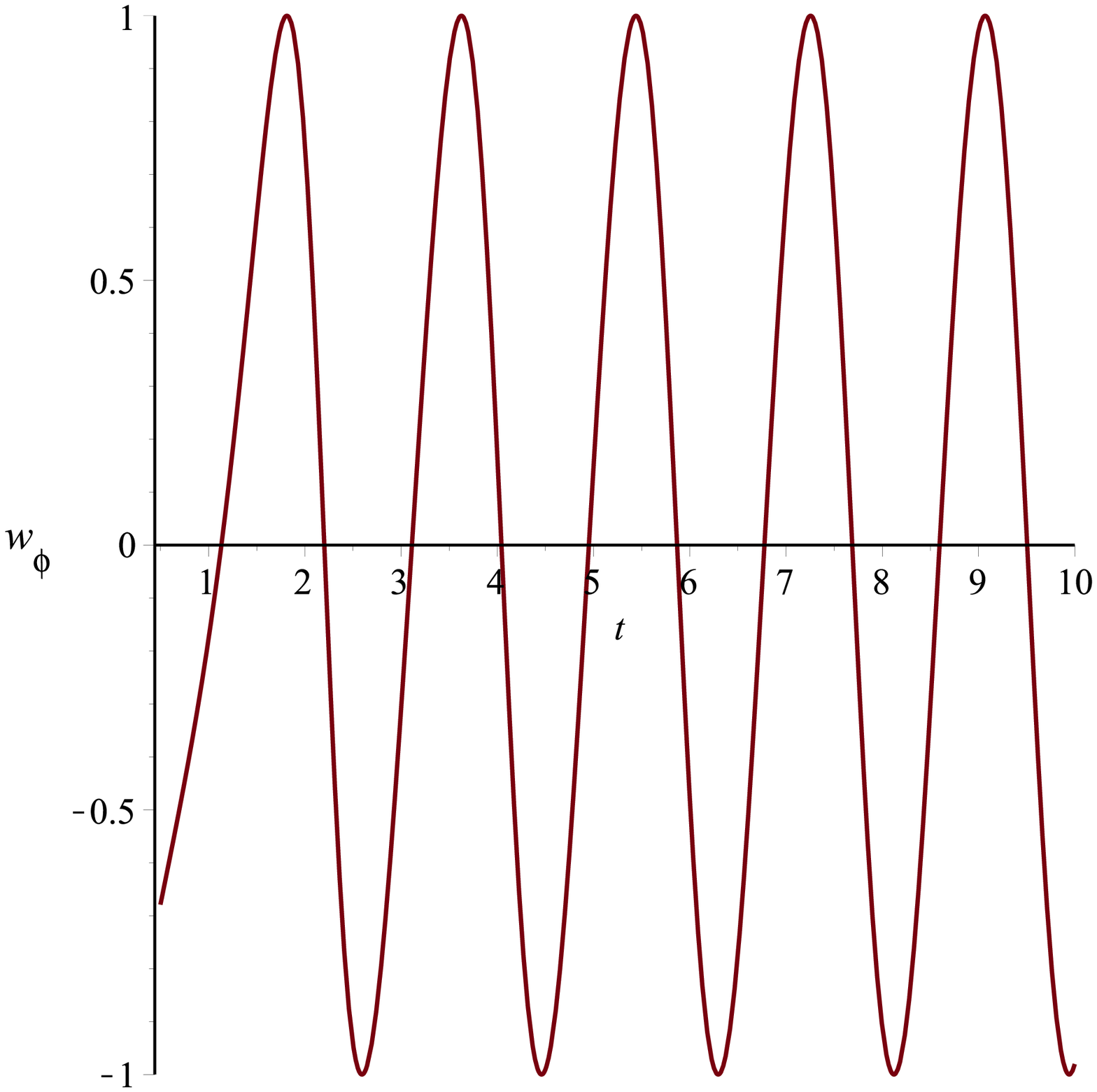}\\
Figure 3: shows the graphical representation of  $w_\phi$ against $t$ for $\lambda_0>0,~\gamma=1$ with $\lambda_0=8,~b_1=1,~p^2=3,~V_0=4$.
\end{minipage}
\end{figure}

$\underline{B.~for~\lambda_0~{>}~0,~\gamma=0}$

\begin{eqnarray}
&a(t)=&\left(\frac{3A_0u_0}{2}\right)^{\frac{2}{3}}\left[t^2-\frac{1}{l^2}sin^2(lt)\right]^\frac{1}{3},
\nonumber
\\
&\phi(t)=&exp\left[\frac{1}{p}(tanh^{-1}[\frac{sin (lt)}{(lt)}-b_1)\right].
\end{eqnarray}
\begin{figure}
\begin{minipage}{0.4\textwidth}
\includegraphics[width=1.0\textwidth]{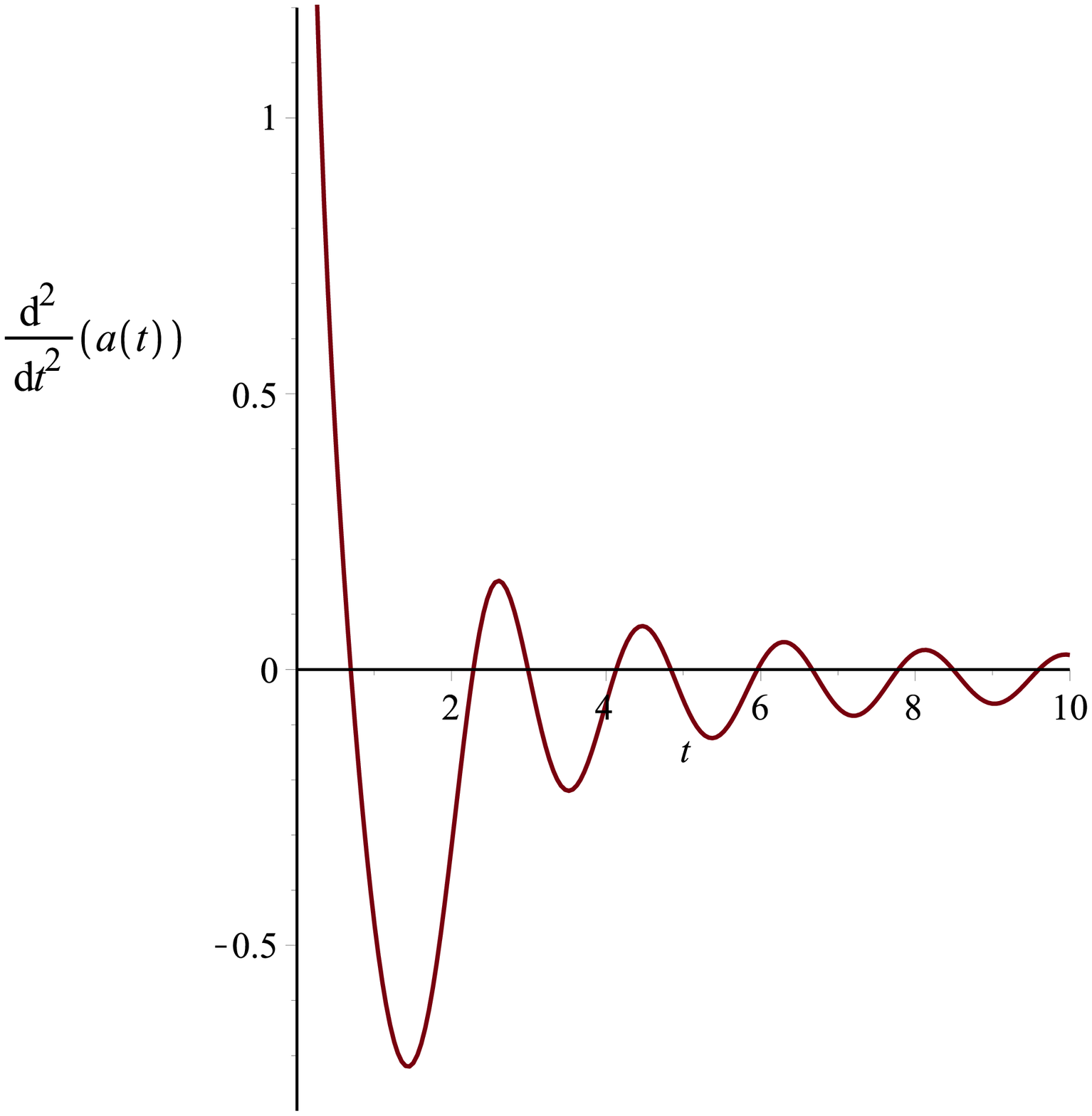}\\
Figure 4: corresponds the graphical representation of  $\ddot{a}(t)$ against $t$ for $\lambda_0>0,~\gamma=0$ with $A_0=1,~u_0=1,~l^2=3,~m^2=\frac{1}{3}$.
\end{minipage}
\begin{minipage}{0.4\textwidth}
\includegraphics[width=1.0\textwidth]{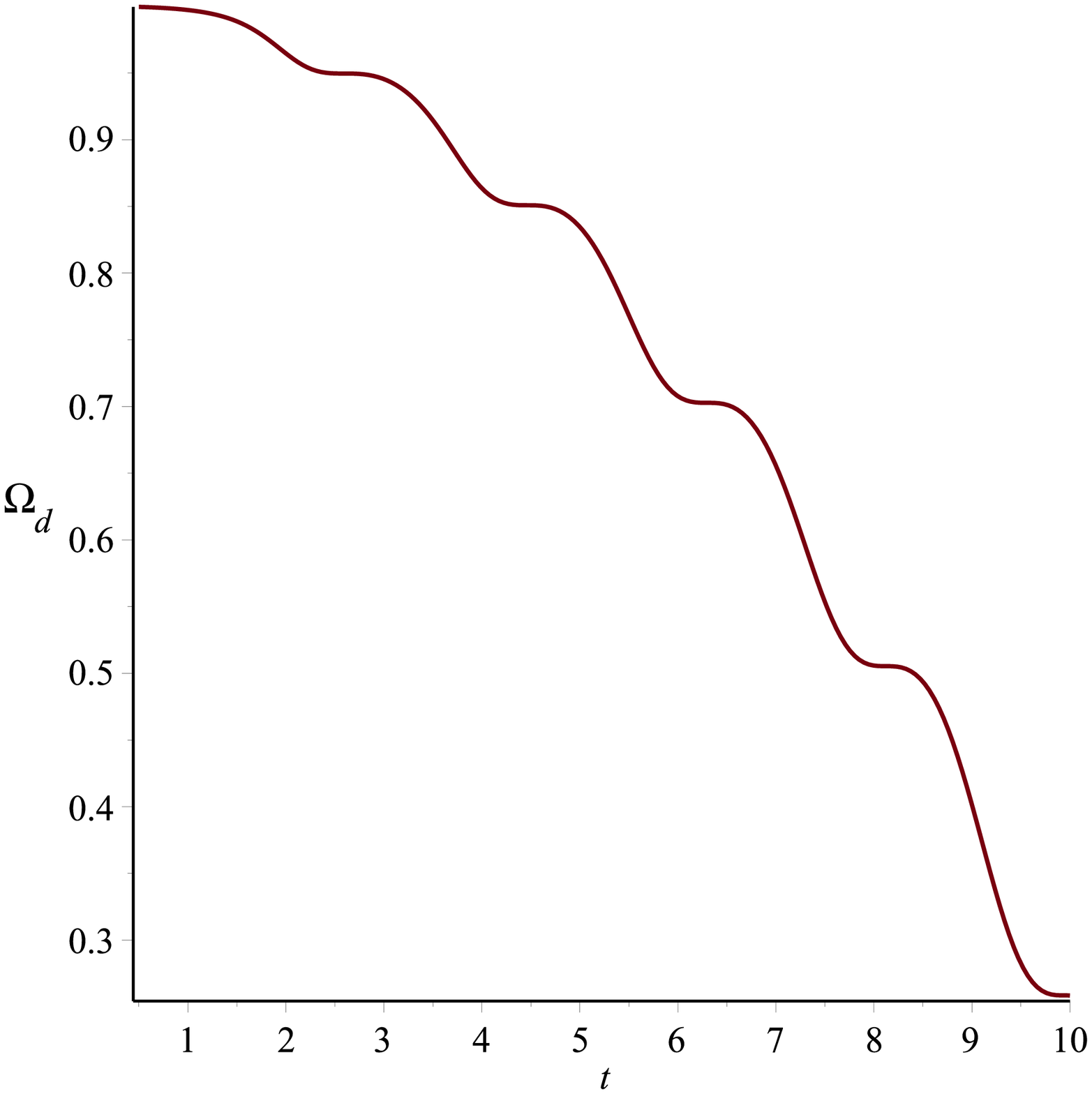}
Figure 5: represents $\Omega_d$ against $t$ for $\lambda_0>0,~\gamma=0$ with $A_0=1,~u_0=1,~l^2=3,~m^2=\frac{1}{3}$.
\end{minipage}
\end{figure}
\begin{figure}
\begin{minipage}{0.4\textwidth}
\includegraphics[width=1.0\textwidth]{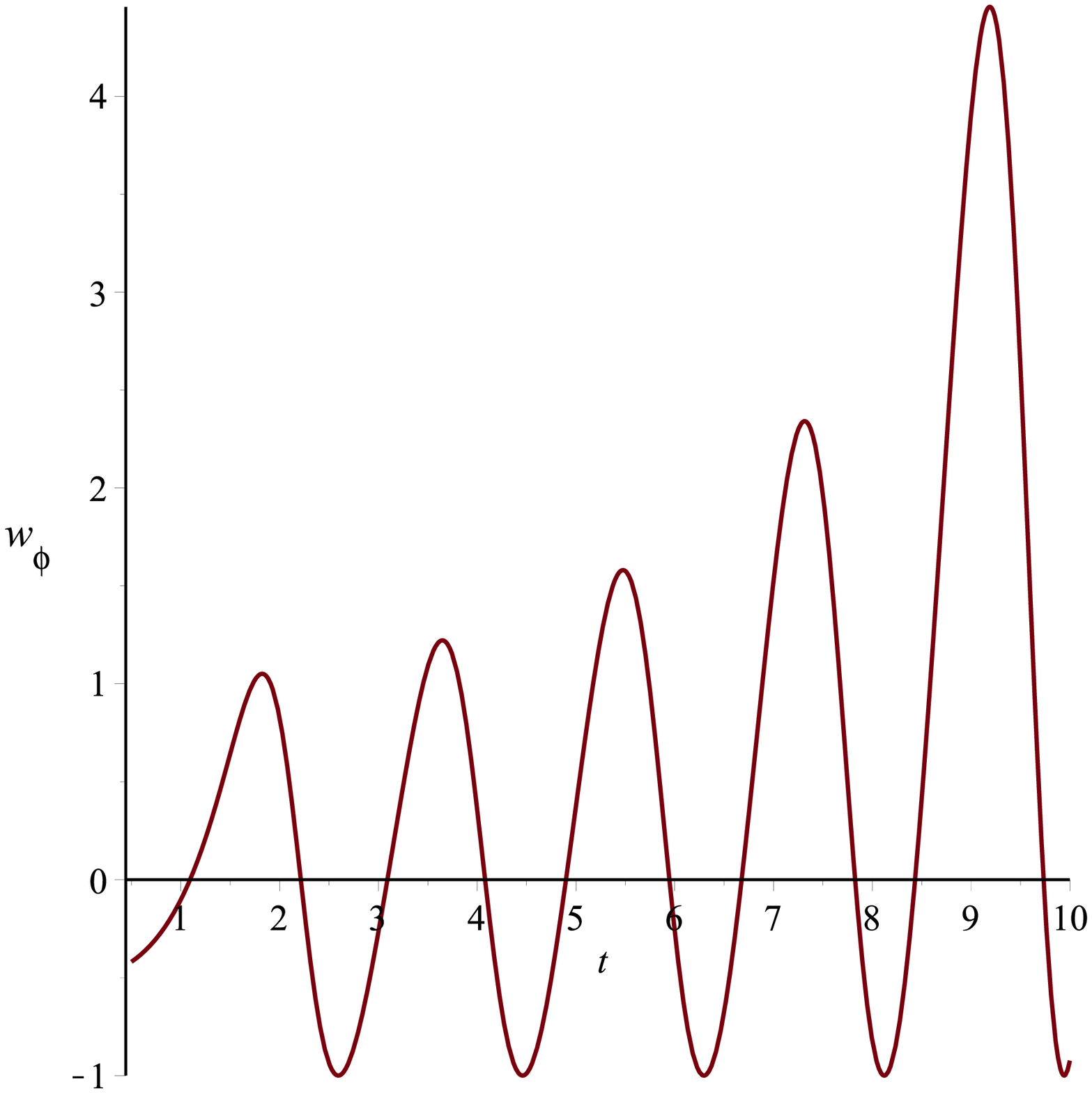}\\
Figure 6: shows the variation of  $w_\phi$ against $t$ for $\lambda_0>0,~\gamma=0$ with $\lambda_0=8,~b_1=1,~p^2=3,~V_0=4$.
\end{minipage}
\end{figure}

$\underline{C.~for~\lambda_0~{<}~0,~\gamma=1}$

\begin{equation}
a(t)=\left\{
\begin{array}{ll}
\left(\frac{3A_0}{2}\right)^{\frac{2}{3}}\left[u_0^2t^2+\frac{(u_0^2-m^2)}{l^2}cosh^2(lt)\right]^{\frac{1}{3}}\\
\left(\frac{3A_0}{2}\right)^{\frac{2}{3}}\left[u_0^2t^2+\frac{(m^2-u_0^2)}{l^2}sinh^2(lt)\right]^{\frac{1}{3}}
\end{array}
\right\},
\end{equation}
and
\begin{equation}
\phi(t)=\left\{
\begin{array}{ll}
exp\left[\frac{1}{p}(tan^{-1}[\frac{\sqrt{u_0^2-m^2}}{lu_0t}cosh(lt)]-b_1)\right]\\
exp\left[\frac{1}{p}(tan^{-1}[\frac{\sqrt{m^2-u_0^2}}{lu_0t}sinh(lt)]-b_1)\right]
\end{array}
\right\},
\end{equation}
according as $u_0~{>}~m~or~u_0~{<}~m.$\\
\begin{figure}
\begin{minipage}{0.4\textwidth}
\includegraphics[width=1.0\textwidth]{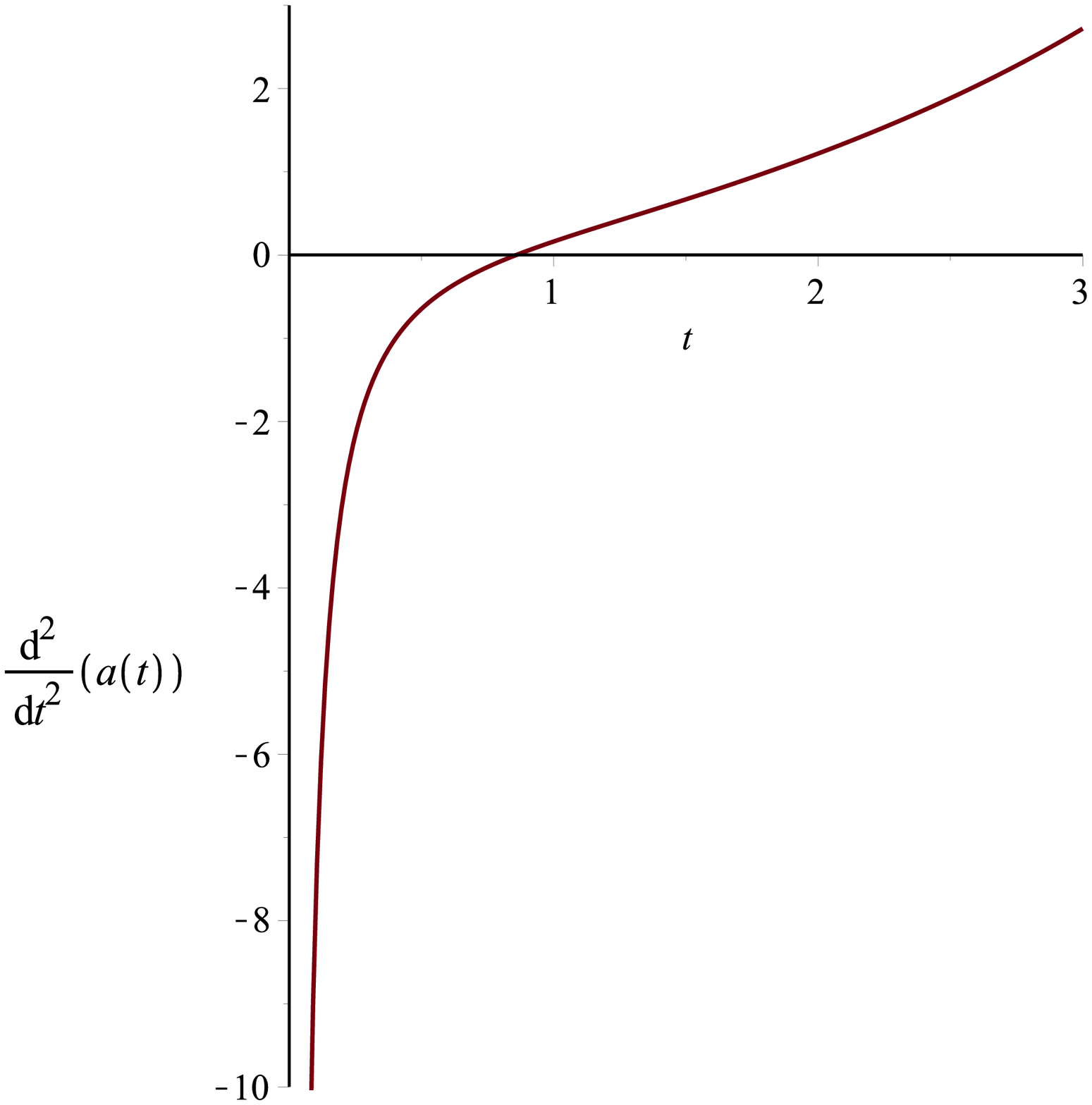}\\
Figure 7: exhibits the variation of  $\ddot{a}(t)$ against $t$ for $\lambda_0<0,~\gamma=1$ with $A_0=1,~u_0=1,~l^2=1,~m^2=2$.
\end{minipage}
\begin{minipage}{0.3\textwidth}
\includegraphics[width=1.0\textwidth]{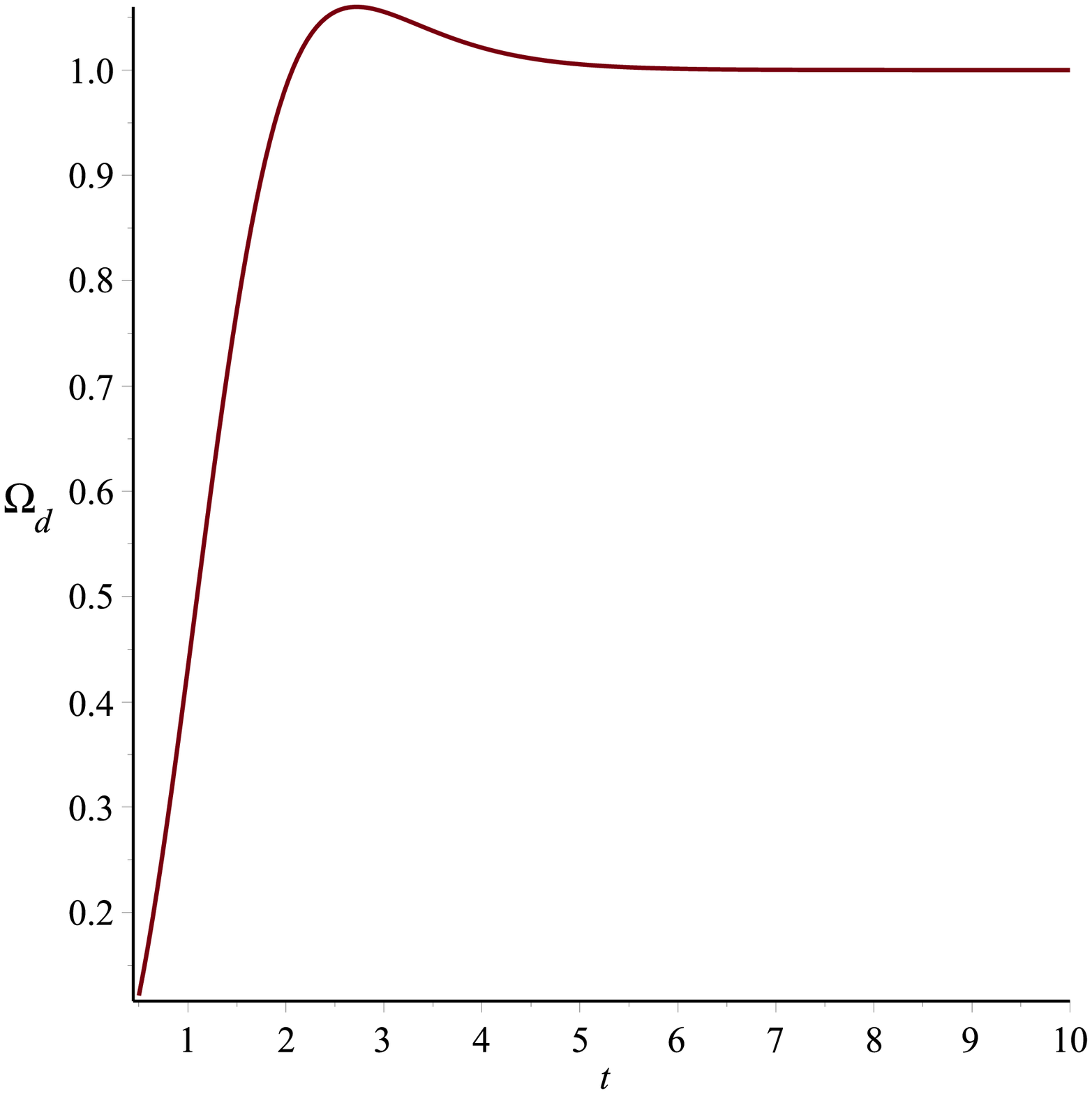}\\
Figure 8: shows the variation of  $\Omega_d$ against $t$ for $\lambda_0<0,~\gamma=1$ with $A_0=1,~u_0=1,~l^2=1,~m^2=2$.
\end{minipage}
\end{figure}
\begin{figure}
\begin{minipage}{0.4\textwidth}
\includegraphics[width=1.0\textwidth]{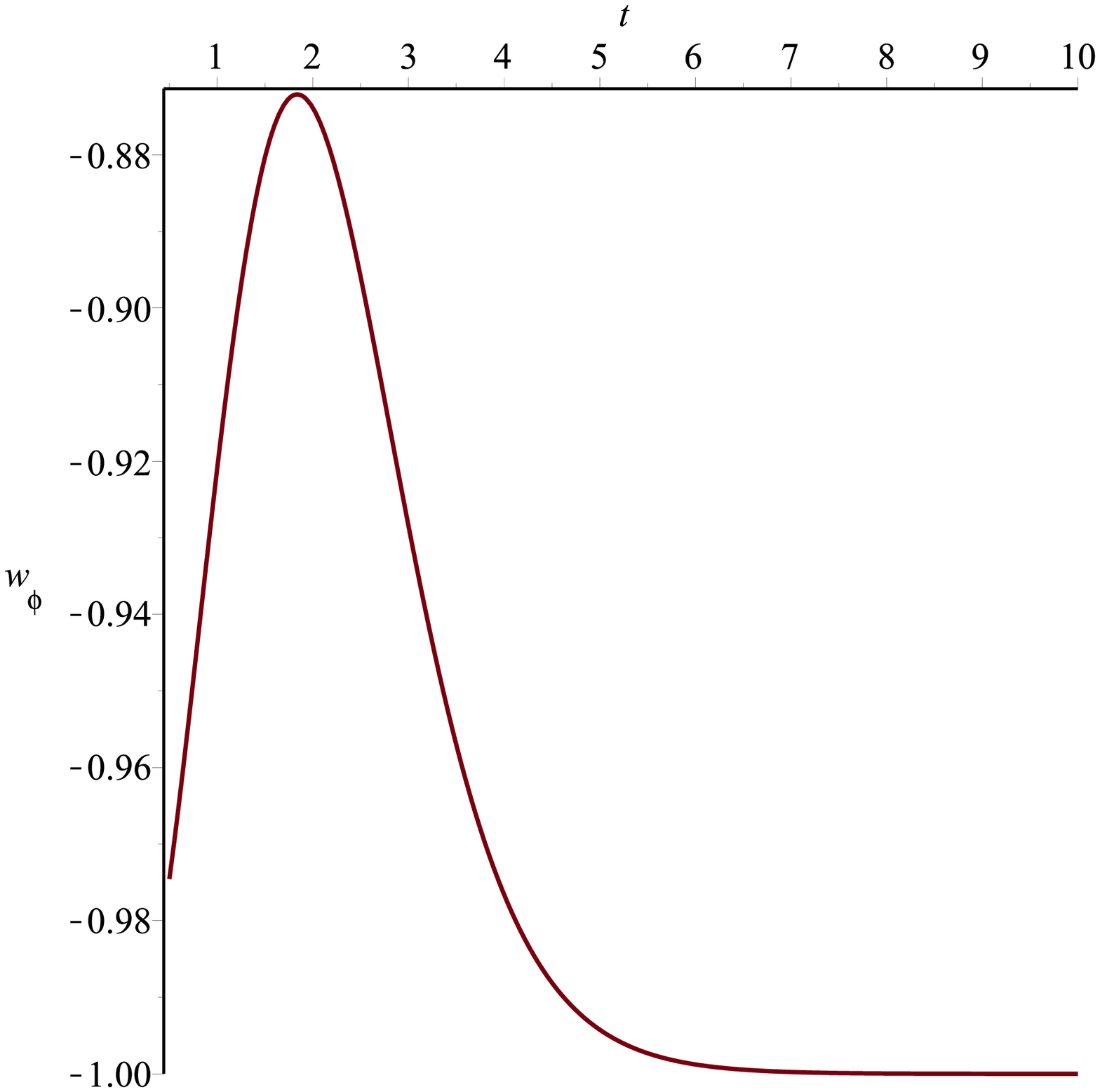}\\
Figure 9: represents  $w_\phi$ against $t$ for $\lambda_0<0,~\gamma=1$ with $\lambda_0=\frac{8}{3},~b_1=\frac{1}{\sqrt{3}},~p^2=1,~V_0=\frac{4}{3}$.
\end{minipage}
\end{figure}

$\underline{D.~for~\lambda_0~{<}~0,~\gamma=0}$

\begin{eqnarray}
&a(t)=&\left(\frac{3A_0u_0}{2}\right)^{\frac{2}{3}}\left[t^2-\frac{1}{l^2}cosh^2(lt)\right]^\frac{1}{3},
\nonumber
\\
&\phi(t)=&exp\left[\frac{1}{p}(tanh^{-1}[\frac{cosh (lt)}{(lt)}-b_1)\right].
\end{eqnarray}

If we now impose the initial big-bang singularity at $t=0$, i.e., $a(0)=0$ then we have the following restrictions:\\
$(i)~u_0{>}m~~for~\lambda_0{>}0~,~(II)~u_0{<}m~~for~\lambda_0{<}0$ and case D is not possible. Thus for $\lambda_0{>}0$ it is possible to have both $\gamma=0~and~1$ while for $\lambda_0{<}0$ only the dark matter in the form of dust is permissible. It should be noted that due to complicated form of the symmetry condition (46), it is not possible to find the cosmological solution corresponding to it. Further, if we set the present time $t_0=1$, i.e., we use the time scale in which the age if the universe (not known in principle) is chosen as unit of time and setting $a_0=a(t_0)=1$ we have

\begin{equation}
B_0^2=\left\{
\begin{array}{lll}
\frac{1}{3}\left(\frac{\frac{4}{3}-\kappa^2 \rho_0 \frac{sin^2l}{l^2}}{1-\frac{sin^2l}{l^2}}\right) & \mbox{ for } \lambda_0>0,\gamma=1\\
\frac{\frac{4}{9}}{1-\frac{sin^2l}{l^2}} & \mbox{ for } \lambda_0>0, \gamma=0\\
\frac{1}{3}\left(\frac{\frac{4}{3}+\kappa^2 \rho_0 \frac{sinh^2l}{l^2}}{1+\frac{sinh^2l}{l^2}}\right) & \mbox{ for } \lambda_0<0,\gamma=1\\
\end{array}
\right\},
\end{equation}
with $B_0=A_0u_0$.\\

Thus $B_0$ is completely determined by the present energy density of the DM $(\rho_0)$ and $l~i.e~v_0$, the constant that determines the scale of the potential. Although $v_0$ is not directly measurable ,but its range of variability can be strongly constrained through its relation with the Hubble constant as follows: Due to our choice of time unit, the expansion rate $H(t)$ is dimensionless so that the present value of the Hubble parameter, i.e., $\widetilde{H}=H(t_0)\sim1$ which is not same to the usual $H_0$ both numerically and dimensionally (dimension of $H_0$ is $kms^{-1}Mpc^{-1}$). In fact, they are related by the relation
\begin{equation}
h=9.9\frac{\hat{H_0}}{\tau},
\end{equation}
where $h=\frac{H_0}{100}$ and $\tau$ is the age of the universe in Gy. Now using $\tau=13.73+0.16-0.15,$ [52] we have $h<0.76$ for $\hat{H_0}\approx 1$. For the present model

\begin{equation}
\hat{H_0}=\left\{
\begin{array}{lll}
\frac{3}{2}\left[B_0^2-\frac{B_0^2-\frac{\kappa^2\rho_0}{3}}{2l}sin 2l\right]& \mbox{ for } \lambda_0>0,\gamma=1\\
\frac{3}{2}B_0^2\left(1-\frac{sin2l}{2l}\right) & \mbox{ for } \lambda_0>0, \gamma=0\\
\frac{3}{2}\left[B_0^2-\frac{\frac{\kappa^2\rho_0}{3}-B_0^2}{2l}sinh 2l\right] & \mbox{ for } \lambda_0<0,\gamma=1\\
\end{array}
\right\}.
\end{equation}

Hence one can constrain the range of validity of $v_0$ for the above value of $H_0$. Thus it is possible to express all the basic cosmological parameters in terms of two parameters $v_0$, the scale of the potential and $\rho_0$, the energy density of dark matter at the present epoch. In fact, using the above solutions for $a(t)$ and $\phi(t)$ we can determine the relevant cosmological parameter namely
\begin{equation}
\rho_\phi=\frac{1}{2}\lambda(\phi)\dot{\phi}^2+V(\phi),~~~~p_\phi=\frac{1}{2}\lambda(\phi)\dot{\phi}^2-V(\phi),
\end{equation}
and
\begin{equation}
w_\phi=\frac{\frac{1}{2}\lambda(\phi)\dot{\phi}^2-V(\phi)}{\frac{1}{2}\lambda(\phi)\dot{\phi}^2+V(\phi)},~~\Omega_\phi=\frac{\rho_\phi}{3H^2},
\end{equation}

and their variations with the evolution of the universe has been shown graphically.\\

Among the three sets of solutions, the first two sets namely for $\left(\lambda_0>0,\gamma=1\right)~\textmd{and}~\left(\lambda_0>0,\gamma=0\right)$ there are several transitions from acceleration to deceleration and vice versa (see figures (1) and (4)) while for the third choice there is only one transition from deceleration to acceleration (see figure (7)). It should be noted that the first two solution sets corresponding to normal scalar field while the third choice describes phantom scalar field. From the graphs of the density parameter (i.e., figures 2, 5 and 8) we see that scalar scalar field dominates over the other matter component in the first set but the scalar field contribution gradually diminishes and finally become zero with the evolution of the universe in the second set of solutions. For the third set i.e.for phantom scalar field, the universe is ultimately dominated completely by the phantom field. Finally, figures 3, 6 and 9 show that the equation of state parameter sometimes behaves as normal fluid and sometimes as exotic one for the first two choices and for phantom solution it is always exotic in nature. However, in all the three cases the scalar field extends up to $\Lambda$CDM model.
\section{Summary}
In the present paper, we have studied phantom scalar field cosmology in Einstein gravity using both Lie and Noether symmetries. From Lie symmetry condition, we are able to determine the coupling function to the kinetic term of the phantom field, the potential to the phantom scalar and the equation of state parameter of the matter field. Also this symmetry gives a very simple solution to the phantom scalar field. Subsequently, Noether symmetry is imposed to the cosmological problem. Assuming the coupling function as obtained from Lie symmetry, we have obtained a general form of potential function for the scalar field. Also choosing the exponential potential we have derived the coupling function using Noether symmetry and it is found that the nature of the scalar field (i.e., normal or phantom) is characterized by the sign of $\lambda_0$ ($+$ve or $-$ve). Using conserved quantities corresponding to Noether symmetry the Lagrangian of the system can be written in suitable transformed variables of which one is cyclic, so that the field equations are simple to solve. It is important to mentioned that our results include that of [33] in the case of phantom field and the reason for the existence of symmetries of differential equations is independent on the coordinate system, for a discussion see [44] and for an application see [57]. The relevant cosmological parameters are presented graphically for the admissible cosmological solutions in the figures (1--9). It is to be noted that some of the parameters involved in the solutions can be estimated from the data released by PLANCK satellite [53]. In particular the Hubble parameter $\hat{H_0}$ in equation (67) (or  (68)) can be estimated by the PLANCK data so that it is possible to have some estimation of the parameter $A_0$. \\

Finally, it should be noted that the symmetry conditions may behave as identification of the physically relevant model among different similar models and also helps to solve the problem. The imporatance of the application of symmetries in cosmological studies has been showed and for other kind of symmetries such as the symmetries of the Wheeler-DeWitt equation of quantum cosmology, or the application of Killing Tensors in the graviational Lagrangian see Ref. [58].
\section{acknowledgement}
Author SC thanks Inter University Center for Astronomy and Astrophysics (IUCAA), Pune, India for their warm hospitality as a part of the work was done during a visit. Also SC thanks UGC-DRS programme at the Department of Mathematics, Jadavpur University. Author SD thanks Department of Science and Technology (DST), Govt. of India for awarding Inspire research fellowship.

\end{document}